\documentclass[round]{pnastwo}
\usepackage{graphicx}
\usepackage{color}
\usepackage{amssymb}
\usepackage{amsfonts}
\usepackage{amscd}
\usepackage{latexsym}
\usepackage{amsmath}
\usepackage{amsbsy}
\usepackage{amsgen}
\usepackage{pifont}
\usepackage{gensymb}
\usepackage[english]{babel}
\usepackage[utf8]{inputenc}
\usepackage[numbers]{natbib}
{\rule{20pt}{1ex}\endlist}
\let\oldmarginpar\marginpar
\renewcommand\marginpar[1]{\-\oldmarginpar[\raggedleft\footnotesize #1]%
{\raggedright\footnotesize #1}}

\contributor{Submitted to Proceedings
of the National Academy of Sciences of the United States of America}
\url{www.pnas.org/content/early/2016/05/11/1602451113}
\copyrightyear{2016}
\issuedate{Issue Date}
\volume{Volume}
\issuenumber{Issue Number}

\makeatletter                                  
\def\DonormalEndcol{%                              
\ifx\toporbotfloat\xtopfloat%                          
  \ifcaptypefig%                               
  \expandafter\gdef\csname topfloat\the\figandtabnumber\endcsname{%    
  \vbox{\vskip\PushOneColTopFig%                       
  \unvbox\csname figandtabbox\the\loopnum\endcsname%               
  \vskip\abovefigcaptionskip%                          
  \csname caption\the\loopnum\endcsname%                   
  \csname letteredcaption\the\loopnum\endcsname%               
  \csname continuedcaption\the\loopnum\endcsname%              
  \csname letteredcontcaption\the\loopnum\endcsname            
  \ifredefining%                               
  \csname label\the\loopnum\endcsname%                     
  \expandafter\gdef\csname topfloat\the\loopnum\endcsname{}\fi}%       
  \vskip\intextfloatskip%%                         
  \vskip-4pt %% probably an artifact of topskip??              
}%                                     
\else%                                     
  \ifcaptypeplate%                             
  \expandafter\gdef\csname topfloat\the\figandtabnumber\endcsname{%    
  \vbox{\vskip\PushOneColTopFig%                       
  \unvbox\csname figandtabbox\the\loopnum\endcsname            
  \vskip\abovefigcaptionskip                           
  \csname caption\the\loopnum\endcsname                    
  \csname letteredcaption\the\loopnum\endcsname                
  \csname continuedcaption\the\loopnum\endcsname               
  \csname letteredcontcaption\the\loopnum\endcsname            
  \ifredefining                                
  \csname label\the\loopnum\endcsname                      
  \expandafter\gdef\csname topfloat\the\loopnum\endcsname{}\fi}        
  \vskip\intextfloatskip %%                            
  \vskip-4pt %% probably an artifact of topskip??              
}%                                     
\else% table ==>                               
 \expandafter\gdef\csname topfloat\the\figandtabnumber\endcsname{%     
 \vbox{\vskip\PushOneColTopTab %%                      
 \csname caption\the\loopnum\endcsname                     
  \csname letteredcaption\the\loopnum\endcsname                
  \csname continuedcaption\the\loopnum\endcsname               
  \csname letteredcontcaption\the\loopnum\endcsname            
  \vskip\captionskip                               
  \unvbox\csname figandtabbox\the\loopnum\endcsname            
\ifredefining                                  
\csname label\the\loopnum\endcsname                    
\expandafter\gdef\csname topfloat\the\loopnum\endcsname{}\fi           
}\vskip\intextfloatskip %% why don't we need this?             
\vskip-10pt}                                   
\fi\fi%                                    
\else% bottom float                            
\ifcaptypefig                                  
\expandafter\gdef\csname botfloat\the\figandtabnumber\endcsname{%      
\vskip\intextfloatskip                             
\vbox{\unvbox\csname figandtabbox\the\loopnum\endcsname            
\vskip\abovefigcaptionskip                         
  \csname caption\the\loopnum\endcsname                    
  \csname letteredcaption\the\loopnum\endcsname%               
  \csname continuedcaption\the\loopnum\endcsname%              
  \csname letteredcontcaption\the\loopnum\endcsname%               
\vskip\PushOneColBotFig%%                          
\ifredefining%                                 
\csname label\the\loopnum\endcsname                    
\expandafter\gdef\csname botfloat\the\loopnum\endcsname{}\fi}}%        
\else                                      
\ifcaptypeplate                                
\expandafter\gdef\csname botfloat\the\figandtabnumber\endcsname{%      
\vskip\intextfloatskip                             
\vbox{\unvbox\csname figandtabbox\the\loopnum\endcsname            
\vskip\abovefigcaptionskip                         
  \csname caption\the\loopnum\endcsname                    
  \csname letteredcaption\the\loopnum\endcsname%               
  \csname continuedcaption\the\loopnum\endcsname%              
  \csname letteredcontcaption\the\loopnum\endcsname%               
\vskip\PushOneColBotFig%%                          
\ifredefining%                                 
\csname label\the\loopnum\endcsname                    
\expandafter\gdef\csname botfloat\the\loopnum\endcsname{}\fi}}%        
  \else% TABLE                                 
\expandafter\gdef\csname botfloat\the\figandtabnumber\endcsname{%      
  \vskip\intextfloatskip                           
\vbox{\csname caption\the\loopnum\endcsname                
  \csname letteredcaption\the\loopnum\endcsname                
  \csname continuedcaption\the\loopnum\endcsname               
  \csname letteredcontcaption\the\loopnum\endcsname%               
  \vskip.5\intextfloatskip                         
  \unvbox\csname figandtabbox\the\loopnum\endcsname%               
\vskip\PushOneColBotTab                            
\ifredefining%                                 
\csname label\the\loopnum\endcsname                    
\expandafter\gdef\csname botfloat\the\loopnum\endcsname{}\fi}}%        
\fi\fi\fi}                                 
\makeatother                                   

\begin{document}
%
%%%\title{Spider capture silk inspires solid-liquid hybrid fibers}
%%\title{Spider capture silk inspires the engineering of hybrid fibres with mixed solid-liquid mechanical properties}
%\title{Spider capture silk inspires hybrid fibres with mixed solid-liquid mechanical properties}
\title{In-drop capillary spooling of spider capture thread inspires hybrid fibres with mixed solid-liquid mechanical properties}
\author{Hervé Elettro\affil{1}{Sorbonne Universités, UPMC Univ Paris 06, CNRS, UMR 7190 Institut Jean Le Rond d'Alembert, F-75005 Paris, France.}
, Sébastien Neukirch\affil{1}{}, Fritz Vollrath\affil{2}{Oxford Silk Group, Zoology Department, University of Oxford, UK} \and Arnaud Antkowiak\affil{1}{}}
\date{\today}

\contributor{Submitted to Proceedings of the National Academy of Sciences
of the United States of America}

\significancetext{The spiraling capture threads of spider orb webs are covered with thousands of tiny glue droplets whose primary function is to entrap insects. In this paper we demonstrate that the function of the drops goes beyond that of glueing preys for they also play a role in the mechanical properties of these fibres -- usually ascribed solely to the complex molecular architecture of the silk. Indeed each of the droplets can spool and pack the core silk filament, thus keeping the thread and the whole web under tension. We demonstrate that this effect is the result of the interplay between elasticity and capillarity by making a fully artificial drops-on-fibre compound as extensible as capture thread is.}

\maketitle

\begin{article}

\begin{abstract}
An essential element in the web-trap architecture, the capture silk spun by ecribellate orb spiders consists of glue droplets sitting astride a silk filament. Mechanically this thread presents a mixed solid/liquid behaviour unknown to date. Under extension, capture silk behaves as a particularly stretchy solid, owing to its molecular nanosprings, but it totally switches behaviour in compression to now become liquid-like: it shrinks with no apparent limit while exerting a constant tension. Here, we unravel the physics underpinning the unique behaviour of this "liquid wire" and demonstrate that its mechanical response originates in the shape-switching of the silk filament induced by buckling within the droplets. Learning from this natural example of geometry and mechanics, we manufactured novel programmable liquid wires that present novel pathways for the design of new hybrid solid-liquid materials.
\end{abstract}

%\pacs{Valid PACS appear here}

\dropcap{H}ybrids made of different materials often display effective properties far exceeding those of their components \citep{Ashby2003}: zinc-coated steel is both strong and corrosion-resistant, metal foams (hybrids of metal and air) are stiff, light and crushable at the same time, making them perfect candidates to absorb energy in a car crash \citep{Gibson1997,Banhart2002}. Nature also provides many exquisite examples of hybrid design such as the seashell nacre, both stiff and tough thanks to its inner `brick-and-mortar' structure composed of rigid, though brittle, inclusions surrounded by a crack arresting soft organic matrix \citep{Dunlop2010}, % the fibre-reinforced structure of the sea urchin teeth allowing for literally grinding rocks, though both the rock and the tooth are almost entirely composed of calcite \citep{Ma2009} 
or the bamboo stem with its hollow core and honeycomb-shaped cells that maximize the ratio of bending rigidity over weight \citep{Wegst2015}. 
%In these examples, the effective properties of the biological composite as well as its mechanical function invariably arise from the microstructure. 
A most interesting natural hybrid material is the spider's capture thread, which consists of a core filament that supports glue droplets.
Here we report on the arresting mechanical behaviour of this capture thread, that changes from solid-like in extension to liquid-like in compression. We trace this behaviour back to the core filament's buckling inside the droplets. A synthetic version of this natural system then allows us to copy the remarkable properties of spider's capture thread to a novel type of hybrid material.% capable for example of accommodating any extension or compression without sag. 

%\dropcap{M}aterials can conveniently be described through Ashby diagrams (or material properties charts)  representing one property of the material, say thermal conductivity, strength or toughness, versus another one [ref Ashby]. Whatever properties represented, it is remarkable to observe the emergence of clusters in such maps: materials of the same class, for example woods or ceramics, all lie in the same region, leaving large empty zones in the material properties space. Obviously if new materials were to be found in these areas, novel design possibilities (in the broad engineering sense) would be uncovered. Interestingly, one way to access artificially these zones is to design hybrids, ie a combination of materials arranged in some particular geometry [Ref Brechet]. Can such compounds still be termed ``materials''? It is a fact that these composites present effective mechanical properties -- sometimes outperforming the ones of their elementary constituents -- and as such can be introduced in the material properties chart. There are no rules to design effectively a priori an hybrid for it achieves some desired properties

\begin{figure*}[ht]
\begin{center}
\includegraphics[width=183mm]{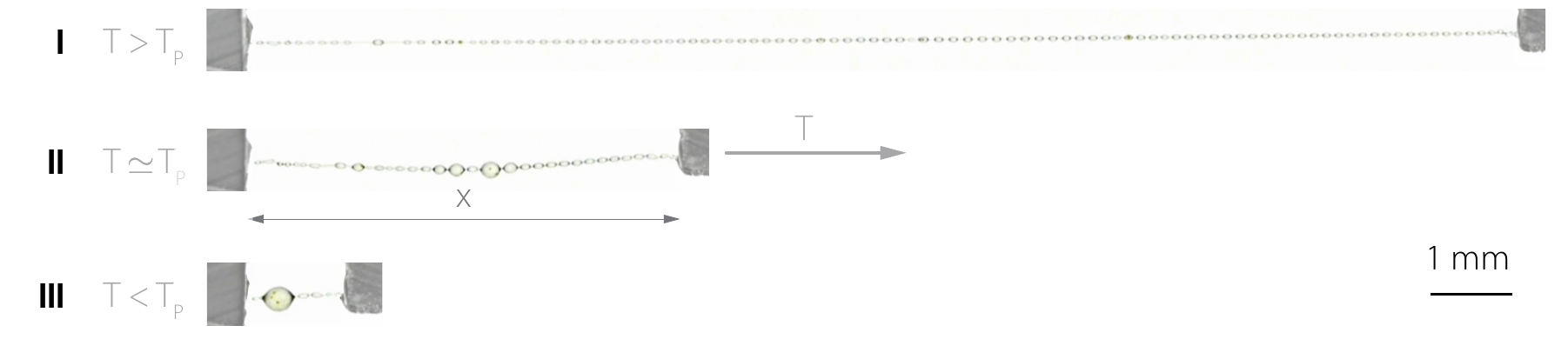}
\caption{\textbf{A liquid-like fibre}. Whether stretched or relaxed, the typical capture silk thread of an araneid orb spider (here \textit{Nephila edulis}) remains taut.
Force monitoring reveals that when subjected to large tension $T$ the fibre behaves like a spring (\textsf{I}).
As $T$ is decreased, a force plateau $T \simeq T_\text P$ is reached, along which the thread adopts a wide range of lengths, just as soap films do (\textsf{II}).
At lower tensions, $T < T_\text{P}$, the thread is totally contracted (\textsf{III}).  See Supplementary Movie S4 for full cycle.}
\label{fig:natural-caliper}
\end{center}
\end{figure*}

Spiders use different kinds of silk to build their webs, and a typical ecribellate orb-web combines dry and smooth radial threads with wet and droplet-covered spiral threads \citep{Foelix2010,Opell2001,Vollrath2006b,Opell2008,Blackledge2009}. The adhesive nature of these droplets enables the spiral capture thread to perform its primary function of catching insect preys \citep{Foelix2010}.
%capture threads form the easily recognizable spirals circling orb-webs.
%The primary function of these dedicated threads (out of seven different spun by araneid spiders) is obviously to endow the web with adhesion. 
%This biological function actually arises from the compound structure of spiral capture threads, made of thin solid flagelliform fibres and gluey liquid aggregate silk droplets 
Apart from being sticky, these capture threads also prove to be particularly resilient to tensile tests: extensive studies on their mechanical behaviour \citep{Denny1976,Foelix2010}  revealed that, when stretched, the thread elongates to three times its web-length without breaking and recoils back with no noticeable hysteresis or sagging when relaxed \citep{Vollrath1989}. 
This stretchiness confers spider silk a strength tenfold that of natural or synthetic rubber \citep{Gosline1984,Gosline1999}.
%Though stretchy, capture silk is nonetheless ten times stronger than natural or synthetic rubber \citep{Gosline1999}.
%
These remarkable extensional properties rely on the macromolecular architecture of capture silk \citep{Becker2003,Blackledge2005}. The ability to cope with stretch is crucial for spider capture threads for it provides their unusually large toughness (energy required for rupture), which in turn allows them to absorb the kinetic energy of incident preys without breaking.
Far less understood is the behaviour of the thread when compressed: unlike any solid fibre that sags or buckles, it keeps taut and self-adapts to compression. \noindent Figure~\ref{fig:natural-caliper} illustrates this singular behaviour, reminiscent of the response of liquid films to compression events: liquid films do not buckle upon squeezing, but rather self-adapt \citep{Gennes2003}. And as for liquid films, self-adaptation for the capture thread is an indication for fibre self-tension.
This liquid-like behaviour in compression suggests that more than merely endowing the web with adhesion, capture silk might well have the additional mechanical function of preserving the web structural integrity. Indeed, without self-adaptation, single sticky strands would touch during relaxation events and thereby irremediably damage the web. With sagging suppressed, the sticky strands are secured apart. 

In the present paper, we investigate and disentangle the mechanism underpinning the unique behaviour of spider capture silk. Based on these insights we design a mechanical hybrid that behaves as a solid when stretched, but as a liquid when compressed. 

\begin{figure}[t]
\begin{center}
\includegraphics[width=89mm]{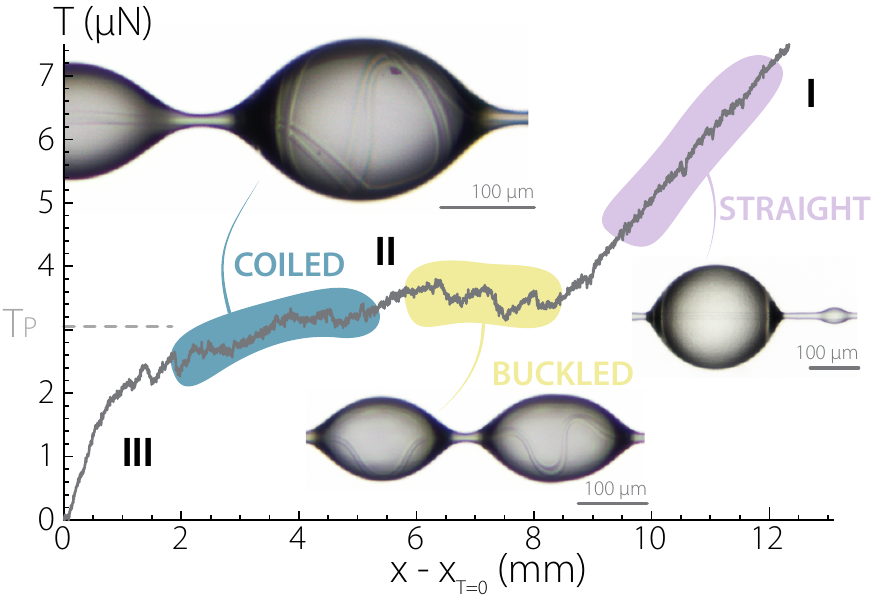}
\caption{\textbf{Shape-induced functionalization.} Quasi-static force measurements on spider capture threads combined with microscopic observations reveal that the core filament coils into the droplets ($\sim$250-300 $\mu$m wide) along a force plateau $T \sim T_\text P$ (liquid-like response). 
%of the minute forces needed to pull a spider capture thread combined with microscopic observations within the glue droplets reveal that the force plateau $T_P$ corresponds to a state where the slack fibre is spooled in the droplets.
%
For larger forces $T>T_\text P$, the fibre straightens and a solid-like behaviour is recovered.
%switched on as soon the fibre becomes straight again.
The particular shape of this force-extension curve can be attributed to a shape-induced functionalization of the fibre by the glue droplets. See also Supplementary Movies S1 and S2.}
\label{fig:natural-jcurve}
\end{center}
\end{figure}

The spectacular macroscopic properties of hybrids often originate in a physical effect that occurs at the micro-structural level (which needs not be molecular, see \textit{e.g.} the buckling of the walls of a unit cell in a cellular solid \citep{Gibson1997,Bertoldi2010}). To investigate the physics of the mechanical hybrid character of spider capture thread, we performed mechanical tests on a single thread alongside microscopic observations of its microstructure. Figure~\ref{fig:natural-jcurve} shows the relaxation of a freshly harvested biological sample. Starting from a stretched state (region \textsf{I}), the force-elongation curve shows that the thread behaves as a regular elastic solid undergoing relaxation: the monitored tension decreases almost linearly with the imposed displacement. In this regime, the capture thread adopts a classic drop-on-straight filament conformation, evocative of unduloidal-shaped drops sitting astride textile fibres \citep{Adam1937}, glass filaments \citep{Quere1999}, mammalian hairs \citep{Carroll1989}, or feathers \citep{Duprat2012}. But as relaxation proceeds further, the mechanical behaviour of the capture thread switches from solid to liquid. This sudden change can be read directly from the mechanical testing: in region \textsf{II}, the recorded tension becomes virtually independent of the imposed displacement. This plateau tension is the typical signature of the response of liquid or soap films to tensile or compressive sollicitations. Strikingly, this behavioural change coincides with a sharp modification of the micro-structure: while the overall composite remains taut, the core filament now buckles within each glue droplet. At even higher compressions, spools of slack filament form within the drops and keep on accumulating until eventually the overall tension falls (region \textsf{III}). Such spools have previously been observed in samples of post-mortem capture threads, but the physics underlying their formation, and in particular the potential roles of the filament molecular structure or of the glue viscoelasticity in this formation, has remained unclear so far  \citep{Foelix2010}.

\begin{figure}[t]
\begin{center}
\includegraphics[width=89mm]{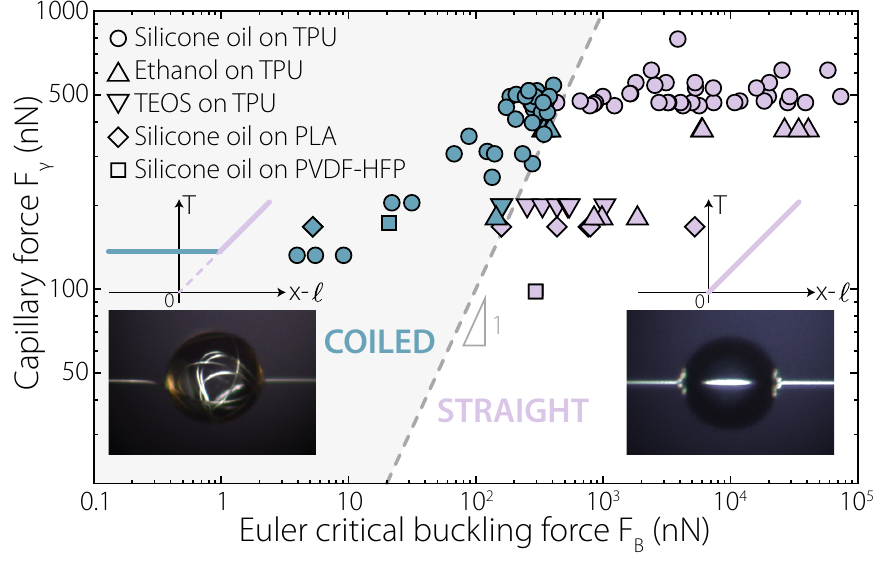}
\caption{\textbf{Spooling activation}. In-drop spooling can also be achieved by synthetic fibres wet by droplets of various newtonian liquids. The phase diagram summarizes experiments performed with different materials and liquids in a quasi-static displacement-controlled setting. Each experiment consists in releasing the external tension on an initially taut system. Spooled or straight filament conformation are then observed within the droplets (blue or purple points respectively). These data demonstrate that the spooling threshold corresponds to a capillarity-induced buckling condition: spooling spontaneously occurs as soon as the capillary force exerted by the drop $F_\gamma$ exceeds the Euler buckling load of the filament $F_B$. Note that, contrary to classic buckling, this spooling  continues to proceed as long as the previous force condition is fulfilled, which suggests a subcritical nature for this elastocapillary instability. The composite overall mechanical response (sketched in insets) also sharply changes past the threshold to exhibit a liquid-like plateau force.}
\label{fig:windlass_activation}
\end{center}
\end{figure}
The coincidence between the change in the mechanical responses of the capture thread at the global scale and the change in the conformations of the core filament at the drop scale is intriguing and requires further investigation. For so, let us consider a composite system consisting of a synthetic core filament and of a liquid droplet, and examine the link between the global mechanical response of the system and the local filament geometry. Specifically we investigate the possibility of a buckling-induced activation of the composite. Surface tension is known to promote buckling \citep{Neu2007,Roman2010}, snapping \citep{Fargette2014}, or wrinkling \citep{Huang2007} of thin lamellar structures.
In the drop-filament composite, and in absence of any external load,  local buckling is initiated when the capillary force developed near each meniscus of a single drop $F_\gamma = 2 \pi h \gamma \cos \theta$ exceeds the Euler buckling load $F_B = k EI / D^2$, where $h$, $\gamma$, $\theta$, $EI$, $D$, and $k$ denote respectively the filament radius, the liquid-air surface tension, the contact angle of the liquid on the filament, the bending stiffness of the core filament, the wet length, and the Euler buckling factor. 
The $h^4$ scaling of the filament's bending stiffness constitutes however a strong restriction for capillarity-induced buckling, typically limiting the manifestation of this phenomenon in filaments in the nm-$\mu$m range -- thereby supporting the observed in-drop buckling of micronic spider capture threads, while explaining why hairs of 80 $\mu$m diameter do not buckle when wet, but rather simply clump \citep{Bico2004}. This fully mechanical scenario, involving capillarity and elasticity as only ingredients, suggests that any drop sitting astride any filament could make it buckle, provided the force condition $F_\gamma > F_B$ is satisfied.
To test this hypothesis, we conducted extensive experiments with various Newtonian liquid drops surrounding synthetic (\textit{i.e.} non-biological) filaments of different diameters and made of diverse materials. Upon release of external tension, we found in-drop elastocapillary buckling to be indeed activated as soon as the capillary force overcomes the Euler buckling load, irrespective of the materials involved, see Fig.~\ref{fig:windlass_activation} and Supplementary Movie S5.
Note that we have here used the value $k=\pi^2$ for the Euler buckling factor, expressing the fact that the fibre can freely rotate at the meniscii (simply supported buckling), see \cite{Elettro2015a}.
Contrary to conventional buckling, past the elastic instability threshold the core filament is not gently deformed but literally spooled and packed within the droplets, although the applied capillary force is constant. This behaviour, along with the localization of the bending deformation, are typical signatures of a subcritical instability.
Furthermore, the global mechanical response of the composite changes instantly as soon as buckling is initiated at the drop scale: under large stretching, the composite behaviour is that of the core filament, but switches to that of a liquid film when compressed past the threshold (see also Supplementary Movie S3). Thus the droplets have the double role of storing the excess thread and putting the whole composite in a state of tension. This behaviour is all the more arresting because real liquid cylinders instantaneously disintegrate due to Rayleigh-Plateau instability, making the liquid-like response of the composite  truly unusual (see Supplementary Movie S1 for an illustration of the composite  in action).

\noindent The geometry of slender elastic objects is known to control their mechanical response \citep{Audoly2010,Lazarus2012}. %(as for \textit{e.g.} springs).
The composite under study here is no exception and we now explain how the in-drop filament geometry leads the thread to inherit the solid core filament mechanical properties when stretched, but the liquid drop properties when compressed. To shed light on this connection between the micro-structure and the global mechanical response,
we consider a simple model where a bendable and stretchable elastic filament supports a liquid drop, the overall system now being  subjected to an external tension~$T$. 
Elastocapillary spooling activation can be described as a phase transition between a wet and coiled phase -- where the filament is entirely packed within the liquid drop -- and a dry and extended phase -- where the filament runs straight outside the drop.
The extended phase is characterized by a stretching modulus $EA$ and a rest length $\ell_\mathrm{e}$. Under an applied tension, its extension is $x_\mathrm{e} = (1+\epsilon_\mathrm{e}) \ell_\mathrm{e}$, where $\epsilon_\mathrm{e}$ is the extensional strain. The strain energy of the phase is then $\frac{1}{2} \, \ell_\mathrm{e} \, EA {\epsilon_\mathrm{e}}^2$, to which we add the solid-air interface energy $2 \pi  h  \gamma_\text{sv} \ell_\mathrm{e}$.
The coiled phase is made up of the drop and the spooled filament inside the drop. The spools certainly adopt a complicated shape and the bending energy of the filament is $\frac{1}{2}  EI \int_0^{\ell_\mathrm{c}} \kappa(s)^2 \, \mathrm{d}s$ where $\kappa$ is the curvature of the filament and $I = \pi h^4/4$. Approximating the drop as spherical and the spools as arcs of circle, we write $\kappa = 2/D$ where $D$ is the diameter of the drop. The bending energy is then $2 \, \ell_\mathrm{c} \, EI / D^2$. We note that in this approximation the extension of the phase $x_\mathrm{c} = D$ is constant. We add the solid-liquid interface energy $2 \pi  h  \gamma_\text{sl} \ell_\mathrm{c}$ (the liquid-air interface energy, a constant, is not included) to obtain the total energy of the system $V=\left(\frac{1}{2} \, EA {\epsilon_\mathrm{e}}^2 + 2 \pi  h  \gamma_\text{sv} \right) \ell_\mathrm{e} + \left(2 \, EI / D^2 + 2 \pi  h  \gamma_\text{sl} \right)  \ell_\mathrm{c}$. 
%As $V$ is a linear function of $ \ell_\mathrm{e}$ and $ \ell_\mathrm{c}$ where the total length  $\ell = \ell_\mathrm{e}+ \ell_\mathrm{c} $ is constant, we have a phase transition problem. 
We replace $ \ell_\mathrm{c}$ and, discarding constant terms, re-write the total energy as $V=\left(\frac{1}{2} \, EA {\epsilon_\mathrm{e}}^2 -2 \,  EI / D^2 + 2 \pi  h  \gamma \cos \theta \right) \ell_\mathrm{e}$. Note that we have used Young-Dupr\'e wetting relation $\gamma_\text{sv}-\gamma_{\text{sl}} = \gamma \cos \theta$, where $\theta$ is the liquid contact angle on the filament and $\gamma$ the liquid-air interface energy per area.
We note that the latent energy cost per unit length $\epsilon_0 =  2 \pi h \gamma \cos \theta - \pi E  h^4 / 2 D^2$ involved in the transformation from the coiled to the extended phase is a typical signature of a first-order phase transition problem. From this expression we readily obtain a condition for spooling to be sustained. Indeed, for the coiled phase to be stable at small forces $\epsilon_0$ has to be positive. This condition can be recast into a condition for the radius, where we recover the fact that only  thin filaments exhibit in-drop spooling:
\begin{equation}
h < \left( 4 \gamma \cos \theta \right)^{1/3} \,  E^{-1/3} \, D^{2/3}
\end{equation}
Introducing the ratio $\rho =  \ell_\mathrm{e} / \ell$, we minimize $V$ under the constraints of fixed extension $x=x_\mathrm{c}+x_\mathrm{e}$, and bounded ratio $0 \leq \rho \leq 1$. In the limit where $D \ll \ell$ and $\epsilon_0 \ll EA$, we find that the system can be entirely in the coiled phase ($\rho = 0$; filament fully packed in the drop) with tension $0 < T < \epsilon_0$, or entirely in the extended phase ($\rho=1$) with tension $T = EA (x/\ell - 1) > \epsilon_0$. A third interesting possibility consists in a mixture of phases $0<\rho<1$. In this latter case, part of the filament is packed in the drop while the outer part is taut, consistent with our observations. As $\rho$ is changed, the tension remains constant to a plateau value $T= T_\text{P}=\epsilon_0$, with
\begin{equation}
T_\text{P} = 2 \pi h \gamma \cos \theta -  \pi E \, h^4 / 2D^2 \label{equa:T-analytic}
\end{equation}
\begin{figure}[t]
\begin{center}
\includegraphics[width=89mm]{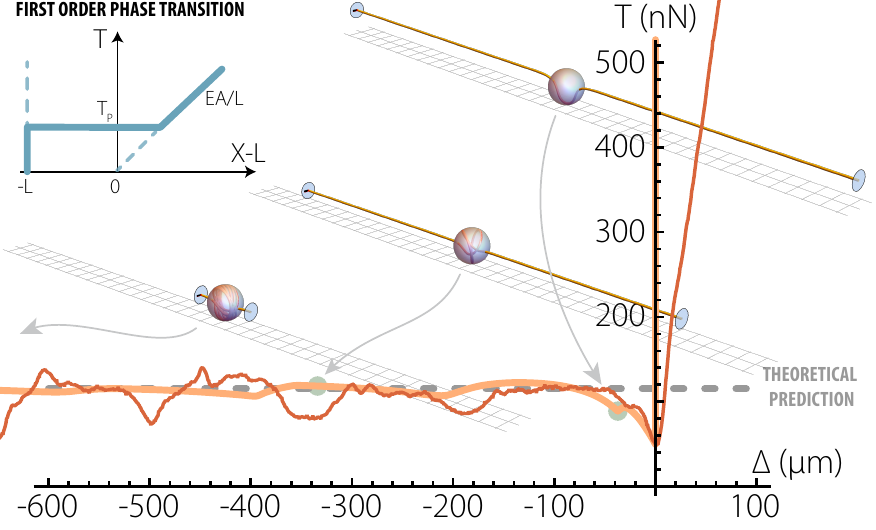}
\caption{\textbf{Structural phase transition and detailed mechanical response.} Comparison between nano-Newton-resolved measurements on a composite polyurethane filament/silicone oil thread (red line),  detailed simulations of an Elastica interacting with a droplet (orange line) and the first-order phase transition model (dashed grey line, full behavior also sketched in inset). Experiments were  performed with a drop of wet length $D = 62 \pm 2 \mu$m and a filament of radius $h = 1 \pm 0.2 \mu$m and Young modulus $E = 17 \pm 3 $MPa. Numerical equilibria are here followed with a continuation procedure, with $F_\gamma = 35 EI / D^2$ and $L = 20 D$. The plateau tension $T_\text{P}$ given by the phase transition model~(\ref{equa:T-analytic}) is here 115 nN.
Beyond the nice overall agreement, the results reveal a difference between the buckling threshold and the plateau tension. This difference points to the subcritical nature of elastocapillary buckling, also evidenced by the sudden localisation of the filament visible in the insets. The numerical simulations allow to capture the fine details in the micro-mechanical response observed in the experiments, resulting in inhomogeneities in the Maxwell plateau.
Sensor drift forced us to adjust the reference level for the experimental measurements, but the level difference between buckling threshold and plateau tension is well recovered. See also Supplementary Movies S3 and S5.}
\label{fig:sma}
\end{center}
\end{figure}
To further explore the mechanical response of the composite system, we also performed detailed numerical computations of equilibrium of an inextensible and flexible elastic filament  \citep{Audoly2010,Antkowiak2011,Elettro2015a}. The filament, held at both extremities with imposed distance $x$, is subjected to attracting meniscus forces $F_{\gamma}$ at entrance and exit of a confining sphere. The loading $(x,T)$ diagram, shown in Fig.~\ref{fig:sma}, reveals inhomogeneities in the Maxwell line \citep{Maxwell1875}. These inhomogeneities are due to fine details in the micro-mechanical response of the system. 
Setting $F_\gamma = 2 \pi h \gamma \cos \theta$, we plot in Fig.~\ref{fig:sma} the phase transition prediction given by Eq.~(\ref{equa:T-analytic}), and we observe a nice agreement not only with the numerical computations, but also with nano-Newton-resolved mechanical testing of synthetic composites (here made of polyurethane filament and silicone oil droplet). We also note that both experiments and numerical simulations exhibit a kink between the two regimes that reveals a difference between the buckling threshold and the plateau tension, as already anticipated by our simple models and by the subcritical nature of the spooling (Fig.~\ref{fig:windlass_activation}).

Unraveling the mechanics of spider capture silk allowed us to design a new type of fully self-assembling hybrid material with unprecedented mechanical function, switching from solid-like in extension to liquid-like in compression.
This bioinspired hybrid can be manufactured with virtually any material, and provides novel functionalities such as fibre spooling or unspooling at the micronic level,  or constant force application for a wide range of extension while preserving tautness and reversibility.
Strikingly, rather than being a failure threshold, buckling in this case proves to be a necessary condition for material activation.
%
%\textbf{faire une estimation du capillary number}
%Thus the extraordinary mechanical behaviour of araneid spider capture silk relies more on the macroscopic composite structure than on the microscopic material properties, underlining that natural selection 
%at the macroscopic scale
%sometimes cleverly tinkers by coopting simple laws of physics  \citep{Jacob1977}. 

%\textsf{\textbf{Text reserve:}} Our results fully support the windlass theory and identify the fundamental mechanism to be dependent on the interplay between elasticity and capillarity.  Moreover, we demonstrate empirically as well as theoretically that this mechanism is generic, \textit{i.e.} not requiring silk proteins in either the filaments nor the droplets as fundamental material components.  This not only elucidates the physics of the system but also opens the way for the design of novel bio-inspired synthetic actuators using the windlass array concept. 
%Our results report on the first in-situ observation of windlass within glue droplets and pinpoint the fundamental mechanism responsible for fibre spooling as an interplay between elasticity and capillarity. We demonstrate that this mechanism is largely material independent by endowing synthetic fibres with spider capture thread properties, hence opening the way for the design of novel bioinspired synthetic actuators.

\begin{materials}
\textbf{Capture silk samples.} Our \textit{Nephila edulis} spider was kept in a $80\times80\times30$ cm vivarium, consisting of wood panels, PMMA windows and artificial plants. The spider was kept at high humidity (above 70\%) and comfortable temperature (above $22^\circ$C)  with a 12/12 hr day/night schedule. The spider was fed crickets and flies three times a week. Sections of web were carefully excised  using a soldering iron for transfer within a rigid frame. To visualise the filament running through each droplet, the humidity was set to 100\% rH for 15 minutes before observation. The humidity was then stepped down to 50\% (observation) rH.\\
\textbf{Artificial samples.} PolyUrethane (TPU, Elastollan 1185A from BASF\textsuperscript{\textregistered}, Young's modulus $E$ = 17 MPa) granules were deposited on a hot plate at $230^\circ$C. After melting, we used a tweezer to pick up a small droplet which  was then stretched quickly while at the same time being released into ambient room temperature. This resulted in the creation of micron-sized, metre-long, soft filaments. The filament was then deposited on the measuring setup as outlined below. A droplet of silicone oil (Rhodorsil\textsuperscript{\textregistered} 47V1000, surface tension $\gamma = 21.1 $mN.m$^{-1}$) was then deposited by gently touching and brushing the filament with a drop hanging from a pipette. PLA (PolyLactic Acid, Young's modulus $E \sim$ 4 GPa in the glassy and $\sim$ 4 MPa in the rubbery state) filaments were processed the same way. PVDF-HFP samples were obtained by electrospinning. A droplet of PVDF-HFP (Young's modulus $E \sim$ 10 MPa) in THF was electrospun at 12kV using a charged syringe tip, at room temperature and relative humidity. We thus obtain polymer cables made of many microfibers. The distribution of fibers radii and corresponding cable bending rigidity is inferred optically using a Leica macroscope. We measured contact angles by superimposing optical images of drops on fibers to corresponding calculated profiles, and found $\theta_Y=23^{\degree}\pm2^{\degree}$ for  TPU/silicone oil, $\theta_Y=19^{\degree}\pm2^{\degree}$ for TPU/ethanol, 
$\theta_Y=31^{\degree}\pm2^{\degree}$ for TPU/TEOS,
$\theta_Y=35^{\degree}\pm2^{\degree}$ for PLA/silicone oil and
$\theta_Y=29^{\degree}\pm2^{\degree}$ for PVDF-HFP/silicone oil. The surface tension of the liquid-air interface was measured to be $\gamma = 22.1 $mN.m$^{-1}$ for ethanol, and
$\gamma = 23.5 $mN.m$^{-1}$ for TEOS.
\textbf{Measurement methods.} Filament samples were transferred to the measuring setup by coiling one end around the tip of a FemtoTools FT-FS1000 (FT-FS100) capacitive deflection force sensor with range 50 nN-1mN (5 nN-100 $\mu$N) and gluing the other end to a glass slide as base. The force sensor was mounted on a linear micro positioner SmarAct SLC-1730 (repetability 0.5 $\mu$m) and measurements are performed through a work station by USB connection. All the tests were performed in stretching at a speed of $25 \mu$m/s, and considering the centimeter size in length of the sample, they can be considered to be quasi-static. The optical setup consisted of a Leica macroscope (VZ85RC) mounted on a micro-step motor and a 3 megapixels Leica DFC-295 camera ($400\times$ zoom, 334 nm/pixel picture resolution) or a D800E Nikon camera with 3 10mm C-mount extension rings (937 nm/pixel video resolution and 374 nm/pixel photo resolution) alternatively. We used a Phlox 50x50 mm backlight, at 60000 lux or alternatively an optical fibre with LED lamp (Moritex MHF-M1002) with circular polarizer. Side views were acquired with a second D800E camera, with a 70mm extension tube and a 100mm macro Zeiss lens (7,27 microns/pixel video resolution).
The force sensor was tared to zero with the fibre compressed slightly more than its slack length, so that it sags, but only minutely, be it for fibres with or without droplet. The measurement of the slack length was performed by pulling on the filament at one end by a few micrometers to straighten the fibre.\\
The TPU fibre diameter measurement was performed using Fiji software. A high-resolution picture of the fibre is analysed using the following steps : the contrast is enhanced up to the point that 0.4\% of the pixels are saturated, then the grey value of the pixels on a line perpendicular to the fibre axis is plotted. The typical curve obtained this way resembles a downward pointing gaussian, thus the diameter of the fibre is extracted as the full width at half minimum of the peak.\\
\textbf{Numerical computations.}
The windlass system is modeled as an elastic filament, obeying Kirchhoff equilibrium equations, in interaction with a sphere. Except at the two `meniscus' points, the filament is prevented from touching or crossing the sphere through a soft-wall barrier potential. The equilibrium of the system is solved using two-points boundary-value problem techniques (shooting method in Mathematica, and collocation method using the Fortran - AUTO code).
\end{materials}

\begin{acknowledgments}
The present work was supported by ANR grants  ANR-09-JCJC-0022-01 and ANR-14-CE07-0023-01, `La Ville de Paris - Programme \'Emergence', Royal Society International Exchanges Scheme 2013/R1 grant IE130506, and the PEPS PTI program from CNRS. Support from the European Research Council (SP2-GA-2008-233409) and the US-AFOSR (FA9550-12-1-0294) is also acknowledged.
We thank Régis Wunenburger for discussions and experimental advices on thread visualization within droplets, Christine Rollard for advices in spider housing and Natacha Krins for electrospinning the PVDF-HFP filaments. We also acknowledge Yves Bréchet for an enlightening discussion on hybrid materials.
\end{acknowledgments}

% \paragraph{$\rhd$ Authors' contribution.}
% A.~A. designed and conducted the experiments, ran the simulations, wrote
% the paper, and contributed to the design of the model.  B.~A. designed
% the model, wrote the simulation code, and wrote the paper.  S.~N.
% contributed to a preliminary form of the model and wrote the paper.
% M.~R. conducted the experiments and contributed to a preliminary form of
% the model.

\bibliographystyle{unsrt}

\begin{thebibliography}{10}

\bibitem{Ashby2003}
M.F. Ashby and Y.J.M. Br{\'e}chet.
\newblock Designing hybrid materials.
\newblock {\em Acta Materialia}, 51(19):5801 -- 5821, 2003.

\bibitem{Gibson1997}
Lorna~J Gibson and Michael~F Ashby.
\newblock {\em Cellular solids: structure and properties}.
\newblock Cambridge University Press, 1997.

\bibitem{Banhart2002}
John Banhart and Denis Weaire.
\newblock On the road again: Metal foams find favor.
\newblock {\em Physics Today}, 55(7):37--42, 2002.

\bibitem{Dunlop2010}
John~W.C. Dunlop and Peter Fratzl.
\newblock Biological composites.
\newblock {\em Annual Review of Materials Research}, 40(1):1--24, 2010.

\bibitem{Wegst2015}
Ulrike G.~K. Wegst, Hao Bai, Eduardo Saiz, Antoni~P. Tomsia, and Robert~O.
  Ritchie.
\newblock Bioinspired structural materials.
\newblock {\em Nature Materials}, 14(1):23--36, 2015.

\bibitem{Foelix2010}
Rainer Foelix.
\newblock {\em Biology of spiders}.
\newblock Oxford University Press, 2010.

\bibitem{Opell2001}
Brent~D. Opell and Jason~E. Bond.
\newblock Changes in the mechanical properties of capture threads and the
  evolution of modern orb-weaving spiders.
\newblock {\em Evolutionary Ecology Research}, 3(5):507--519, 2001.

\bibitem{Vollrath2006b}
Fritz Vollrath.
\newblock Spider silk: Thousands of nano-filaments and dollops of sticky glue.
\newblock {\em Current Biology}, 16(21):R925 -- R927, 2006.

\bibitem{Opell2008}
Brent~D. Opell, Brian~J. Markley, Charles~D. Hannum, and Mary~L. Hendricks.
\newblock The contribution of axial fiber extensibility to the adhesion of
  viscous capture threads spun by orb-weaving spiders.
\newblock {\em Journal of Experimental Biology}, 211(14):2243--2251, 2008.

\bibitem{Blackledge2009}
Todd~A. Blackledge, Nikolaj Scharff, Jonathan~A. Coddington, Tamas Sz{\"u}ts,
  John~W. Wenzel, Cheryl~Y. Hayashi, and Ingi Agnarsson.
\newblock Reconstructing web evolution and spider diversification in the
  molecular era.
\newblock {\em Proceedings of the National Academy of Sciences},
  106(13):5229--5234, 2009.

\bibitem{Denny1976}
Mark Denny.
\newblock The physical properties of spider's silk and their role in the design
  of orb-webs.
\newblock {\em The Journal of Experimental Biology}, 65(2):483--506, 1976.

\bibitem{Vollrath1989}
Fritz Vollrath and Donald~T. Edmonds.
\newblock Modulation of the mechanical properties of spider silk by coating
  with water.
\newblock {\em Nature}, 340:305--307, 1989.

\bibitem{Gosline1984}
John~M. Gosline, Mark~W. Denny, and M.~Edwin DeMont.
\newblock Spider silk as rubber.
\newblock {\em Nature}, 309(5968):551--552, 1984.

\bibitem{Gosline1999}
J.M. Gosline, P.A. Guerette, C.S. Ortlepp, and K.N. Savage.
\newblock The mechanical design of spider silks: from fibroin sequence to
  mechanical function.
\newblock {\em Journal of Experimental Biology}, 202(23):3295--3303, 1999.

\bibitem{Becker2003}
Nathan Becker, Emin Oroudjev, Stephanie Mutz, Jason~P. Cleveland, Paul~K.
  Hansma, Cheryl~Y. Hayashi, Dmitrii~E. Makarov, and Helen~G. Hansma.
\newblock Molecular nanosprings in spider capture-silk threads.
\newblock {\em Nature Materials}, 2(4):278--283, 2003.

\bibitem{Blackledge2005}
Todd~A. Blackledge, Adam~P. Summers, and Cheryl~Y. Hayashi.
\newblock Gumfooted lines in black widow cobwebs and the mechanical properties
  of spider capture silk.
\newblock {\em Zoology}, 108(1):41--46, 2005.

\bibitem{Gennes2003}
P.G. de~Gennes, F.~Brochard-Wyart, and D.~Qu\'er\'e.
\newblock {\em Capillarity and Wetting Phenomena: Drops, Bubbles, Pearls,
  Waves}.
\newblock Springer, 2003.

\bibitem{Bertoldi2010}
K.~Bertoldi, P.~M. Reis, S.~Willshaw, and T.~Mullin.
\newblock Negative poisson's ratio behavior induced by an elastic instability.
\newblock {\em Advanced Materials}, 22(3):361--366, 2010.

\bibitem{Adam1937}
N.~K. Adam.
\newblock Detergent action and its relation to wetting and emulsification.
\newblock {\em Journal of the Society of Dyers and Colourists}, 53(4):121--129,
  1937.

\bibitem{Quere1999}
David Qu{\'e}r{\'e}.
\newblock Fluid coating on a fiber.
\newblock {\em Annual Review of Fluid Mechanics}, 31(1):347--384, 1999.

\bibitem{Carroll1989}
Brendan~Joseph Carroll.
\newblock Droplet formation and contact angles of liquids on mammalian hair
  fibres.
\newblock {\em Journal of the Chemical Society, Faraday Transactions 1:
  Physical Chemistry in Condensed Phases}, 85(11):3853--3860, 1989.

\bibitem{Duprat2012}
C.~Duprat, S.~Protiere, A.~Y. Beebe, and H.~A. Stone.
\newblock Wetting of flexible fibre arrays.
\newblock {\em Nature}, 482(7386):510--513, 2012.

\bibitem{Neu2007}
S{\'e}bastien Neukirch, Beno{\^\i}t Roman, Beno{\^\i}t de~Gaudemaris, and
  Jos{\'e} Bico.
\newblock Piercing a liquid surface with an elastic rod: Buckling under
  capillary forces.
\newblock {\em Journal of the Mechanics and Physics of Solids},
  55(6):1212--1235, 2007.

\bibitem{Roman2010}
B.~Roman and J.~Bico.
\newblock Elasto-capillarity: deforming an elastic structure with a liquid
  droplet.
\newblock {\em Journal of Physics: Condensed Matter}, 22(49):493101, 2010.

\bibitem{Fargette2014}
Aur\'elie Fargette, S\'ebastien Neukirch, and Arnaud Antkowiak.
\newblock Elastocapillary snapping: Capillarity induces snap-through
  instabilities in small elastic beams.
\newblock {\em Phys. Rev. Lett.}, 112:137802, 2014.

\bibitem{Huang2007}
Jiangshui Huang, Megan Juszkiewicz, Wim~H. de~Jeu, Enrique Cerda, Todd Emrick,
  Narayanan Menon, and Thomas~P. Russell.
\newblock Capillary wrinkling of floating thin polymer films.
\newblock {\em Science}, 317(5838):650--653, 2007.

\bibitem{Bico2004}
J.~Bico, B.~Roman, L.~Moulin, and A.~Boudaoud.
\newblock Adhesion: Elastocapillary coalescence in wet hair.
\newblock {\em Nature}, 432(7018):690--690, 2004.

\bibitem{Elettro2015a}
Herv{\'e} Elettro, Fritz Vollrath, Arnaud Antkowiak, and S{\'e}bastien
  Neukirch.
\newblock Coiling of an elastic beam inside a disk: A model for spider-capture
  silk.
\newblock {\em International Journal of Non-Linear Mechanics}, 75:59 -- 66,
  2015.

\bibitem{Audoly2010}
Basile Audoly and Yves Pomeau.
\newblock {\em Elasticity and Geometry: From hair curls to the non-linear
  response of shells}.
\newblock Oxford University Press, 2010.

\bibitem{Lazarus2012}
A.~Lazarus, H.~C.~B. Florijn, and P.~M. Reis.
\newblock Geometry-induced rigidity in nonspherical pressurized elastic shells.
\newblock {\em Phys. Rev. Lett.}, 109:144301, 2012.

\bibitem{Antkowiak2011}
A.~Antkowiak, B.~Audoly, C.~Josserand, S.~Neukirch, and M.~Rivetti.
\newblock Instant fabrication and selection of folded structures using drop
  impact.
\newblock {\em Proc. Natl Acad. Sci. U.S.A.}, 108(26):10400--10404, 2011.

\bibitem{Maxwell1875}
James~Clerk Maxwell.
\newblock On the dynamical evidence of the molecular constitution of bodies.
\newblock {\em Nature}, 11:357--359, 1875.

\end{thebibliography}

\end{article}
\end{document}